\documentclass[twocolumn]{pasj01}

\RequirePackage{mathpazo}
\RequirePackage[T1]{fontenc}

\let\leftroot\relax
\let\uproot\relax

\usepackage{amsmath}

\def\commenta{$^*$}
\def\commentb{$^\dagger$}

\newcounter{author}
\def\authorcount#1#2{\refstepcounter{author}\label{#1}
                     \altaffiltext{\ref{#1}}{#2}}

\Received{$\langle$reception date$\rangle$}
\Accepted{$\langle$acception date$\rangle$}
\Published{$\langle$publication date$\rangle$}

\begin{document}
\SetRunningHead{Y. Tampo et al.}{V627 Peg 2021 superoutburst}

\title{2021 superoutburst of WZ Sge-type dwarf nova V627 Pegasi lacks an early superhump phase}
 
\author{
        Yusuke~\textsc{Tampo}\altaffilmark{\ref{affil:Kyoto}*},
         Taichi~\textsc{Kato}\altaffilmark{\ref{affil:Kyoto}},
         Naoto~\textsc{Kojiguchi}\altaffilmark{\ref{affil:Kyoto}},
        Sergey~Yu.~\textsc{Shugarov}\altaffilmark{\ref{affil:shu}}, 
        Hiroshi~\textsc{Itoh}\altaffilmark{\ref{affil:Ioh}},
        Katsura~\textsc{Matsumoto}\altaffilmark{\ref{affil:oku}}, 
        Momoka~\textsc{Nakagawa}\altaffilmark{\ref{affil:oku}}, 
        Yukitaka~\textsc{Nishida}\altaffilmark{\ref{affil:oku}}, 
        Michael~\textsc{Richmond}\altaffilmark{\ref{affil:rit}},
        Masaaki~\textsc{Shibata}\altaffilmark{\ref{affil:Kyoto}},
        Junpei~\textsc{Ito}\altaffilmark{\ref{affil:Kyoto}}, %
        Gulchehra~\textsc{Kokhirova}\altaffilmark{\ref{affil:SNG}},
        Firuza~\textsc{Rakhmatullaeva}\altaffilmark{\ref{affil:SNG}},
        Tam\'as~\textsc{Tordai}\altaffilmark{\ref{affil:Trt}},
        Seiichiro~\textsc{Kiyota}\altaffilmark{\ref{affil:kis}}, 
        Javier~\textsc{Ruiz}\altaffilmark{\ref{affil:rui1}}$^,$\altaffilmark{\ref{affil:rui2}}$^,$\altaffilmark{\ref{affil:rui3}}
        Pavol~A.~\textsc{Dubovsky}\altaffilmark{\ref{affil:Vih}}, 
        Tom\'a\v{s}~\textsc{Medulka}\altaffilmark{\ref{affil:Vih}}, %
        Elena~P.~\textsc{Pavlenko}\altaffilmark{\ref{affil:CRI}},
        Oksana~I.~\textsc{Antonyuk}\altaffilmark{\ref{affil:CRI}}, 
        Aleksei~A.~\textsc{Sosnovskij}\altaffilmark{\ref{affil:CRI}}, 
        Aleksei~V.~\textsc{Baklanov}\altaffilmark{\ref{affil:CRI}}, 
        Viktoriia~\textsc{Krushevska}\altaffilmark{\ref{affil:shu}}$^,$\altaffilmark{\ref{affil:shu2}}, 
        Tonny~\textsc{Vanmunster}\altaffilmark{\ref{affil:Van1}}$^,$\altaffilmark{\ref{affil:Van2}},
         Stephen~M.~\textsc{Brincat}\altaffilmark{\ref{affil:BSM}},
         Karol~\textsc{Petrik}\altaffilmark{\ref{affil:Hlo}},
        Charles~\textsc{Galdies}\altaffilmark{\ref{affil:gch}},
        Franz-Josef~\textsc{Hambsch}\altaffilmark{\ref{affil:ham1}}$^,$\altaffilmark{\ref{affil:ham2}}$^,$\altaffilmark{\ref{affil:dfs}}, 
        Yutaka~\textsc{MAEDA}\altaffilmark{\ref{affil:mdy}}, 
        and 
         Daisaku~\textsc{Nogami}\altaffilmark{\ref{affil:Kyoto}}
}

\authorcount{affil:Kyoto}{
     Department of Astronomy, Kyoto University, Kitashirakawa-Oiwake-cho, Sakyo-ku, Kyoto 606-8502, Japan}
\email{$^*$tampo@kusastro.kyoto-u.ac.jp}

\authorcount{affil:shu}{
Astronomical Institute of the Slovak Academy of Sciences, 05960 Tatransk\'a Lomnica, The Slovak Republic}

\authorcount{affil:Ioh}{
     Variable Star Observers League in Japan (VSOLJ),
     1001-105 Nishiterakata, Hachioji, Tokyo 192-0153, Japan}

\authorcount{affil:oku}{
Osaka Kyoiku University, 4-698-1 Asahigaoka, Osaka 582-8582, Japan}

\authorcount{affil:rit}{
    Physics Department, Rochester Institute of Technology, Rochester, New York 14623, USA}
 
\authorcount{affil:SNG}{
    Institute of Astrophysics, National Academy of Sciences of Tajikistan, Dushanbe, 734042, Republic of Tajikistan}
 
\authorcount{affil:Trt}{
     Polaris Observatory, Hungarian Astronomical Association,
     Laborc utca 2/c, 1037 Budapest, Hungary}

\authorcount{affil:kis}{
Variable Star Observers League in Japan (VSOLJ), 7-1 Kitahatsutomi, Kamagaya, Chiba 273-0126, Japan}

\authorcount{affil:rui1}{
    Observatorio de Cantabria, Ctra. de Rocamundo s/n, Valderredible, 39220, Cantabria, Spain.}
\authorcount{affil:rui2}{
    Instituto de Fisica de Cantabria (CSIC-UC), Avda. Los Castros s/n, 39005, Santander, Spain}
\authorcount{affil:rui3}{
    Agrupacion Astronomica Cantabra, Apartado 573, 39080, Santander, Spain}

\authorcount{affil:Vih}{
     Vihorlat Observatory, Mierova 4, 06601 Humenne, Slovakia}

\authorcount{affil:CRI}{
    Federal State Budget Scientific Institution "Crimean Astrophysical
    Observatory", Nauchny, 298409, Republic of Crimea}

\authorcount{affil:shu2}{
    Main astronomical observatory of National academy of sciences of Ukraine, 
    27 Akademika Zabolotnoho St., 03143,Kyiv, Ukraine}

\authorcount{affil:Van1}{
    Center for Backyard Astrophycis Belgium, 
    Walhostraat 1a, B-3401 Landen, Belgium}

\authorcount{affil:Van2}{
    Center for Backyard Astrophycis Extremadura, 
    e-EyE Astronomical Complex, 
    ES-06340 Fregenal de la Sierra, Spain}

\authorcount{affil:BSM}{
    Flarestar Observatory, San Gwann SGN 3160, Malta}

\authorcount{affil:Hlo}{
    M. R. Stefanik Observatory and Planetarium, Sladkovicova 41, 92001 Hlohovec, Slovakia}

\authorcount{affil:gch}{
    Institute of Earth Systems, University of Malta, Malta}
    
\authorcount{affil:ham1}{
    Groupe Européen d’Observations Stellaires (GEOS), 23 Parc de Levesville, 28300 Bailleau l’Evêque, France}
\authorcount{affil:ham2}{
    Bundesdeutsche Arbeitsgemeinschaft für Veränderliche Sterne (BAV), Munsterdamm 90, 12169 Berlin, Germany}

\authorcount{affil:dfs}{
    Vereniging Voor Sterrenkunde (VVS), Oostmeers 122 C, 8000 Brugge, Belgium}

\authorcount{affil:mdy}{
Variable Star Observers League in Japan (VSOLJ), Kaminishiyama-machi 12-2, Nagasaki, Nagasaki, Japan}


\KeyWords{accretion, accretion disk --- novae, cataclysmic variables --- stars: dwarf novae --- stars :individual (V627 Pegasi)}

\maketitle

\begin{abstract}

Superoutbursts in WZ Sge-type dwarf novae (DNe) are characterized by both early superhumps and ordinary superhumps originating from the 2:1 and 3:1 resonances, respectively.
However, some WZ Sge-type DNe show a superoutburst lacking early superhumps;
it is not well established how these differ from superoutbursts with an early superhump phase.
We report time-resolved photometric observations of the WZ Sge-type DN V627 Peg during its 2021 superoutburst.
The detection of ordinary superhumps before the superoutburst peak highlights that this 2021 superoutburst of V627 Peg,
like that {in} 2014, did not feature an early superhump phase.
The duration of stage B superhumps was slightly longer in the 2010 superoutburst accompanying early superhumps than that in the 2014 and 2021 superoutbursts which lacked early superhumps.
This result suggests that an accretion disk experiencing the 2:1 resonance may have a larger mass at the inner part of the disk and hence take more time for the inner disk to become eccentric.
The presence of a precursor outburst in the 2021 superoutburst suggests that the maximum disk radius should be smaller than that of the 2014 superoutburst, even though the duration of quiescence was longer than that before the 2021 superoutburst.
This could be accomplished if the 2021 superoutburst was triggered as an inside-out outburst or if the mass transfer rate in quiescence changes by a factor of two, suggesting that the outburst mechanism and quiescence state of WZ Sge-type DNe may have more variety than ever thought.

\end{abstract}


\section{Introduction}
\label{sec:1}

Cataclysmic variables (CVs) are close binary systems made up of an accreting primary white dwarf (WD) and a secondary low-mass star [see \citet{war95book, hel01book} for a general review of CVs].
Dwarf novae (DNe) are a subclass of CVs which possess an  accretion disk and show recurrent outbursts. 
The mechanism of DN outbursts is explained by the thermal instability model in an accretion disk \citep{osa74DNmodel, hos79DImodel, mey81DNoutburst}.

SU UMa-type DNe are {characterized} by a superoutburst, featuring {larger outburst amplitude, longer outburst duration, and, the most importantly,} small variations {in light curve} with a period slightly longer than the orbital period, called superhumps. 
{The mechanism of superoutbursts and  {superhumps}  is widely accepted in the thermal-tidal instability (TTI) model; an accretion disk reaches the 3:1 resonance radius and becomes eccentric, leading to more effective tidal dissipation and brightening \citep{whi88tidal, osa89suuma, hir90SHexcess}.}
{If a mass ratio of a system is below $\sim$0.3, the tidal truncation radius becomes larger than the 3:1 resonance radius and hence superoutbursts and superhumps are observed.
We note that this picture has been challenged by \citet{neu16htcas, neu20HTCasoutburstdopmap, ama21ezlyn} on the basis of the large disk in quiescence inferred by spectroscopic observations.
We discuss this matter in Appendix 1.}
{In SU UMa-type DNe, a precursor outburst is often observed, which is a normal outburst triggering the following superoutburst.
\citet{osa03DNoutburst} discussed that a precursor outburst is observed when the accretion disk reaches the 3:1 resonance radius but the cooling wave starts to propagate before the eccentricity grows enough to trigger a superoutburst. 
The period of superhumps systematically changes throughout a superoutburst: an early stage with a longer and stable period (stage A), a middle stage with $P_{\rm dot} = \dot{P}/P > 0$ (stage B), and a later stage with a shorter period (stage C) superhumps \citep{Pdot}.
The stage A superhumps are considered as a growing phase of the tidal resonance at the 3:1 resonance radius, whereas the stage B superhumps correspond to the inward propagation of the eccentricity in a disk \citep{kat13qfromstageA, nii21a18ey}.
Ordinary superhumps usually show redder color in superhump peaks, and those in stage B show stronger color variation \citep{how02wxcet, mat09v455and, iso15ezlyn, nak13j0120, neu17j1222, ima18HVVirJ0120, ima18j1740}.
The redder color in superhump peaks is because of the strong tidal dissipation at the 3:1 resonance radius, which is located in the outer and cooler part of an accretion disk \citep{bru96oycareclipsemapping}.
}

Some SU UMa-type DNe with lowest mass ratios are classified as WZ Sge-type DNe (see \cite{kat15wzsge}).
When its disk reaches the 2:1 resonance radius, a spiral pattern emerges in the disk, and double-peaked early superhumps are observed in their light curves before the appearance of ordinary superhumps \citep{lin79lowqdisk,lub91SHa, osa02wzsgehump, uem12ESHrecon}.
The period of early superhumps is almost identical to the orbital period of the system (e.g., \cite{ish02wzsgeletter}).
{The mass ratio of WZ Sge-type DNe are typically below 0.1 \citep{kat15wzsge}, while there are some exceptions \citep{wak17asassn16eg}.
The theoretical expectation of the maximum mass ratio of WZ Sge-type DNe is 0.08 according to \citet{osa02wzsgehump}, comparing the 2:1 resonance radius and the angular momentum stored in the accretion disk during quiescence, rather than with the tidal truncation radius.
This is justified by the weak tidal force from the secondary in a low mass ratio system and the sudden expansion of a disk \citep{osa02wzsgehump}.}

While  almost all superoutbursts from WZ Sge-type DNe feature an early superhump phase (WZ Sge-type superoutburst), a handful number of WZ Sge-type DNe show  superoutbursts lacking an early superhump phase (SU UMa-type superoutburst, 
e.g., V627 Peg; \cite{Pdot7}, 
AL Com; \cite{kim16alcom}, 
EG Cnc; \cite{kim21EGCnc}, 
V3101 Cyg; \cite{tam20j2104}, 
MASTER OT J094759.83+061044.4; vsnet-alert
25401\footnote{http://ooruri.kusastro.kyoto-u.ac.jp/mailarchive/vsnet-alert/25401}, 
25403\footnote{http://ooruri.kusastro.kyoto-u.ac.jp/mailarchive/vsnet-alert/25403},
25410\footnote{http://ooruri.kusastro.kyoto-u.ac.jp/mailarchive/vsnet-alert/25410},
25425\footnote{http://ooruri.kusastro.kyoto-u.ac.jp/mailarchive/vsnet-alert/25425},
and possibly AT 2021iko; \cite{sor21AT2020iko}).
\citet{kim16alcom, kim21EGCnc, sor21AT2020iko} interpreted the difference between SU UMa-type and WZ Sge-type superoutbursts as a dependence {on} the mass of the disk at the onset of outbursts and whether the disk radius reaches the 2:1 resonance radius.
In V3101 Cyg, the superoutbursts lacking early superhumps were observed during the rebrightening phase, indicating that the initial mass of the disk was less than that of the main superoutburst with an early superhump phase \citep{tam20j2104}.

In the cases of well-observed WZ Sge-type superoutbursts, one observes a very fast rise to the peak, reaching the peak magnitude typically less than two days (e.g., WZ Sge; \cite{ish02wzsgeletter}, GW Lib; \cite{waa07gwlibiauc, tem07cbet922gwlib}), which is regarded as evidence of outside-in outbursts in WZ Sge-type DNe.
This fact tightly constrains the disk structure in quiescence of WZ Sge-type DNe to prevent it from triggering an inside-out outburst, suggesting either a very low viscous parameter $\alpha \sim 0.001$ in quiescence \citep{sma93wzsge,osa95wzsge} or the truncation of the inner disk by a magnetized WD or by the evaporation effect \citep{las95wzsge, war96wzsge, mat07wzsgepropeller, kuu11wzsge}.
However, we lack solid observational results of the disk evolution
and still seek better evidence for outside-in outbursts in WZ Sge-type DNe.

V627 Peg (=OT J213806.6+261957) was discovered in 2010 \citep{yam10j2138cbet2273, nak10j2138cbet2275}, and spectroscopically confirmed as a DN outburst by \citet{ara10j2138cbet2274,gra10j2138cbet2275, tov10j2138cbet2283}.
Three outbursts of V627 Peg have been reported in literature so far: in 1942; \citet{hud10j2138atel2619}, in 2010; \citet{pdot2, cho12j2138, zem13j2138, mit14j2138}, and in 2014; \citet{Pdot7}.
Although the outburst profile of the 1942 outburst is unknown, those of the 2010 and 2014 superoutbursts showed large differences.
The 2010 superoutburst accompanied possible early superhumps \citep{pdot2, cho12j2138}.
However, the 2014 super-outburst was clearly fainter at the peak than the 2010 superoutburst was, and ordinary superhumps were detected just 1.5 d after the outburst detection. 
These constraints strongly indicated that there was no early superhump phase in the 2014 superoutburst \citep{Pdot7}. 
Therefore, V627 Peg is one of the best studied DN showing both WZ Sge-type and SU UMa-type superoutbursts.
The periods of possible early superhumps and growing (stage A) superhumps are 0.054523 d and 0.05663(4) d \citep{pdot2, Pdot7}, respectively, yielding a mass ratio of $q = 0.12(2)$ by the method proposed by \citet{kat13qfromstageA, kat22updatedSHAmethod}.
Based on this large mass ratio for WZ Sge-type DNe, as well as the relatively large $P_{\rm dot}$, \citet{Pdot7} proposed that V627 Peg may be an SU UMa-type DN rather than a WZ Sge-type DN.
The distance of V672 Peg is estimated to be 98.9(4) pc in Gaia EDR3 \citep{gaiaedr3, Bai21GaiaEDR3distance}, and the galactic reddening toward the direction of V627 Peg at 100 pc is E($g-r$) = 0.00(1) \citep{gre19dustextinction}.
Therefore, we did not consider the galactic extinction in our analysis.

In this paper, we present our time-series and multi-color photometric observations of V627 Peg during its 2021 superoutburst.
Section \ref{sec:2} presents an overview of our observations of the V627 Peg 2021 superoutburst,  and Section \ref{sec:3} shows the results of our analysis. 
We discuss the uniqueness and the possible nature of V627 Peg in Section \ref{sec:4} and summarize our work in Section \ref{sec:5}.

\section{Observations and Analysis}
\label{sec:2}

Our time-resolved CCD photometric observations of V627 Peg were carried out by the Variable Star Network (VSNET) collaboration \citep{VSNET}. 
The instruments are summarized in table E1 \footnote{Table E1 is available only on the online edition as Supporting Information. }, and the logs of the photometric observations are listed in table E2 \footnote{Table E2 is available only on the online edition as Supporting Information. }. 
All the observation epochs in this paper are described in Barycentric Julian Date (BJD). 
Our photometric data include multi-band and unfiltered data, and the zero point of the unfiltered data were adjusted to  the $V$ band observations {performed} by T. Tordai.
{For our filtered observations in the $U, B, V, Rc, Ic, g, r$, and $i$ bands, the magnitude calibration was performed using all or a part of the comparison stars same as \citet{cho12j2138}. }
We also obtained photometric data from ASAS-SN Sky Patrol \citep{ASASSN, koc17ASASSNLC} and the Zwicky Transient Facility (ZTF; \cite{ZTF}) alert broker Lasair \citep{lasair}  to examine the global light curve profiles.
Note that we used the apparent magnitude of Lasair with careful checking of the original images.
These survey data were not included in our period analysis.
We note that there is a nearby star ($G = 15.006(3)$ mag with a separation from V627 Peg of 2.559 arcsec in Gaia EDR3; \cite{gaiaedr3}), and hence survey data taken during quiescence are not reliable due to contamination.

The phase dispersion minimization (PDM; \cite{PDM}) method was applied for period analysis in this paper. 
The 1$\sigma$ errors for the PDM analysis was determined following \citet{fer89error, pdot2}.
Before period analysis, the global trend of the light curve was removed by subtracting a smoothed light curve obtained using locally weighted polynomial regression (LOWESS: \cite{LOWESS}). 
Observed-minus-calculated ($O - C$) diagrams are presented for visualizing superhump period variations, which are sensitive to slight changes of the superhump period.  
In this paper, we use 0.055120 d for the calculated values ($C$), following the example of \citet{pdot2}.

\section{Results}
\label{sec:3}

\subsection{Overall light curve during the 2021 superoutburst}
\label{sec:3.1}

\begin{figure}[tbp]
 \begin{center}
  \includegraphics[width=80mm]{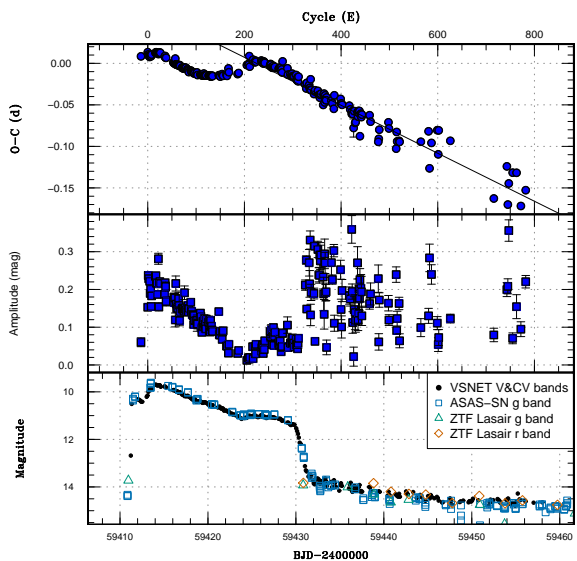}
 \end{center}
 \caption{Top panel : the $O - C$ diagram of superhump periods in the V627 Peg 2021 superoutburst.
 0.055120 d was used for $C$ \citep{Pdot}. 
 The solid line in the panel represents the stage C superhump period of 0.05483 d.
 Middle panel : the evolution of the superhump amplitudes in the magnitude scale. 
 Bottom panel : the light curve of V627 Peg during the 2021 superoutburst.
 Black filled circles, blue open squares, green open triangles, and orange open diamonds represent the data of VSNET collaboration \citep{VSNET} {in $V$ and $CV$ bands}, ASAS-SN $g$ band \citep{ASASSN}, ZTF Lasair $g$ band and $r$ band \citep{lasair}, respectively.}
 \label{fig:2021OC}
\end{figure}

The bottom panel of figure \ref{fig:2021OC} shows the global light curve of V627 Peg during the 2021 superoutburst in the $V$ and $CV$ bands.
The 2021 superoutburst was first reported by Y. Maeda on 15.727 July 2021 = BJD 2459411.230
(vsnet-alert 26059\footnote{http://ooruri.kusastro.kyoto-u.ac.jp/mailarchive/vsnet-alert/26059}).
After an initial brightening at a rate of $\sim8$ mag /d,
this 2021 superoutburst showed an apparent precursor outburst with a peak mag of $\sim 10.3$  from BJD 2459411 to 2459413.
While most of WZ Sge-type DNe do not show any precursor outbursts \citep{kat15wzsge}, some superoutbursts from WZ Sge-type DNe lacking the early superhump phase do show a precursor outburst (e.g., AL Com 2015 superoutburst; \cite{kim16alcom}, EG Cnc 2018 superoutburst; \cite{kim21EGCnc}).
Following the precursor, the outburst reached its optical peak around BJD 2459414.
Its peak magnitude is $\sim$9.6 mag in {the} $V$ band, which is fainter than the 2010 {superoutburst} and is almost equal to the 2014 superoutburst (also see {section \ref{sec:4.1}}).
After a smooth decline from the outburst peak {with a decay rate of $\sim$0.15 mag d$^{-1}$}, {the outburst stopped its decline, slightly brightened, and stayed around 11.0 mag} during BJD 2459424 - 2459430, which we denote as a plateau phase in this paper.
A similar  phase was  observed in its 2010 and 2014 superoutbursts as well, confirming that this is an inherent phenomenon in V627 Peg \citep{pdot2, cho12j2138, Pdot7}.
However, the duration of the plateau phase was the shortest in the 2014 superoutburst and is almost the same in the 2010 and 2021 superoutbursts (also see {section \ref{sec:4.1}}).
{Similar behavior in outburst light curves is observed in some short orbital period and long outburst cycle SU UMa-type DNe as discussed in \citet{kat03hodel, bab00v1028cyg}.
On the other hand, we did not observe such a phase in ordinary WZ Sge-type superoutbursts \citep{kat03hodel}. }
Following a rapid decay phase from the plateau phase, V627 Peg showed a steady decline and no rebrightenings were recorded.

\subsection{superhump profile and color}
\label{sec:3.2}

\begin{figure*}[tbp]
 \begin{center}
  \includegraphics[width=160mm]{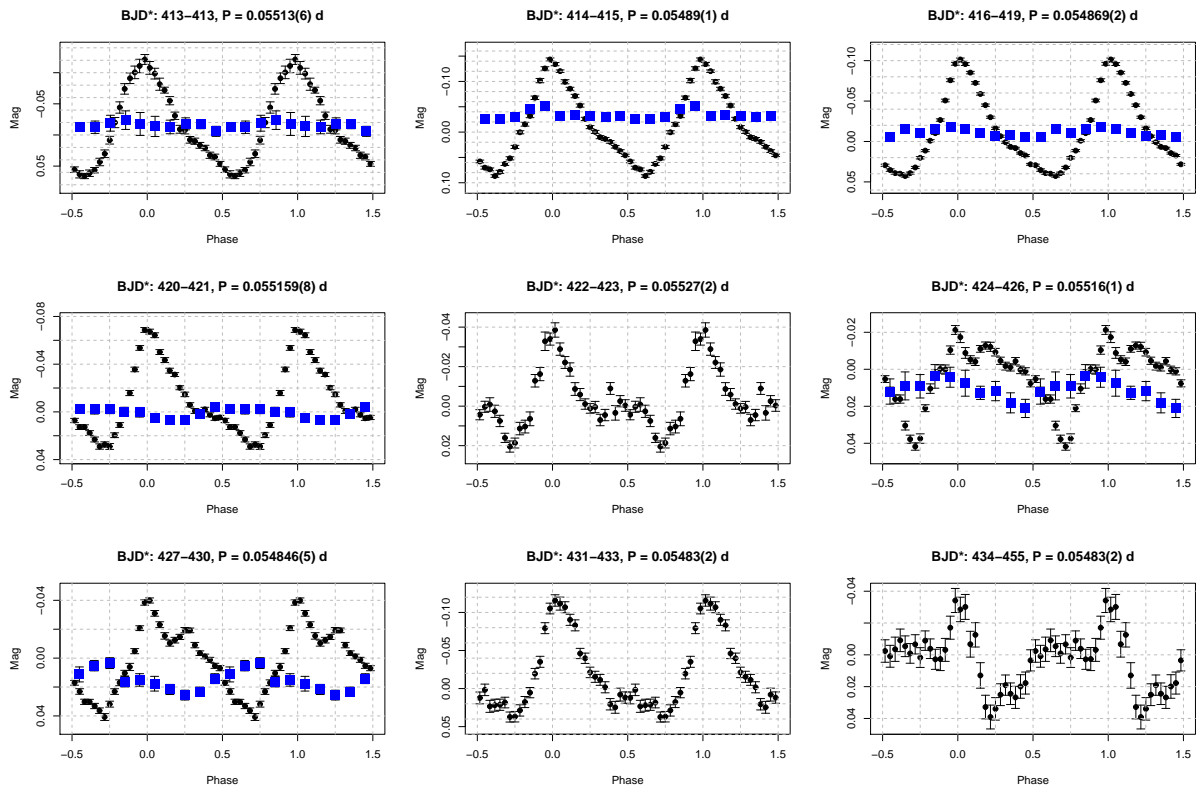}
 \end{center}
 \caption{Phase-averaged profiles during the superoutburst on each epoch.
 The superhump period was obtained through PDM analysis.
 "BJD*" denotes BJD - 2459000.
 Black circles and blue squares represent the $V$ band magnitude and the $V-Rc$ color.}
 \label{fig:dailyPDM}
\end{figure*}

The $O - C$ diagram of the V627 Peg superoutburst in 2021 is presented in the top panel of figure \ref{fig:2021OC} using a period of 0.055120 d as $C$.
The evolution of superhump amplitudes in a magnitude scale is shown in the middle panel of figure $\ref{fig:2021OC}$. 
The {times} of superhump maxima are presented in table E3 \footnote{Table E3 is available only on the online edition as Supporting Information.}.
In figure \ref{fig:dailyPDM}, the phase-averaged superhump profiles are shown in a set of date bins during the superoutburst.
Superhumps were observed since BJD 2459413, which corresponds to the rise after the precursor outburst and before the superoutburst peak epoch.
Therefore, it is most likely that the 2021 superoutburst did not feature any early superhumps.
The superhump periods are 0.05513(6) d on BJD 2459413 and 0.05489(1) d on BJD 2459414 - 2459415, which is clearly shorter than the stage  A superhump period during its 2014 superoutburst (0.05663(4) d; \cite{Pdot7}).
Judging from the period evolution in the $O-C$ diagram and their large superhump amplitudes, we suggest that these superhumps in the 2021 superoutburst are part of stage B superhumps rather than stage A superhumps.
After BJD 2459416, V627 Peg entered a clear stage B superhump phase.
The stage transition from stage B to stage C superhumps occurred around BJD 2459426, at which time V627 Peg was just entering the plateau phase.
After BJD 2459426, stage C superhumps with a stable period of 0.05483(2) were observed (figure \ref{fig:2021OC}), which is consistent with the evolution of the 2010 superoutburst \citep{pdot2}, although daily variations were dominated by flickering. 
We note that our time-resolved observations on BJD 2459412 showed some variation, although no periodicity was detected in our analysis.

After V627 Peg entered the plateau phase, the superhump profiles became double-peaked (figure \ref{fig:dailyPDM}). 
Moreover, during BJD 2459424-2459426, the superhumps changed into flat-top profiles.
This unique profile of superhumps was also observed in the plateau phase of its 2010 superoutburst \citep{cho12j2138, zem13j2138}. 
\citet{zem13j2138} interpreted this flat-top profile as a result of the disk reaching the tidal truncation radius and effective tidal dissipation there.
After the stage transition from stage B to stage C superhumps,
the superhumps returned to top-sharp profiles.
Finally, a double-peaked profile was again observed in the post-superoutburst stage, which is a common feature in the post-superoutburst phase of WZ Sge-type DNe \citep{kat15wzsge}.

The $V-Rc$ color of the superhump profiles is also presented in figure \ref{fig:dailyPDM}.
Our results on BJD 2459413 - 2459415 show almost no color variation.
After BJD 2459416, the superhumps showed the reddest color around the superhump peaks, which is consistent with known superoutbursts in literature \citep{how02wxcet, mat09v455and, iso15ezlyn, nak13j0120, neu17j1222, ima18HVVirJ0120, ima18j1740}.
In the plateau phase (BJD 2459424 - 2459426), even though the superhumps presented a flat-top profile, the superhumps showed a color dependence similar to that of stage B superhumps: bluest around the superhump minimum.
Therefore, the light-emitting region of these humps during the plateau phase is probably located in the outer disk, which is consistent with the interpretation of \citet{zem13j2138}.

\subsection{color evolution}
\label{sec:3.3}

\begin{figure*}[tbp]
 \begin{center}
  \includegraphics[width=80mm]{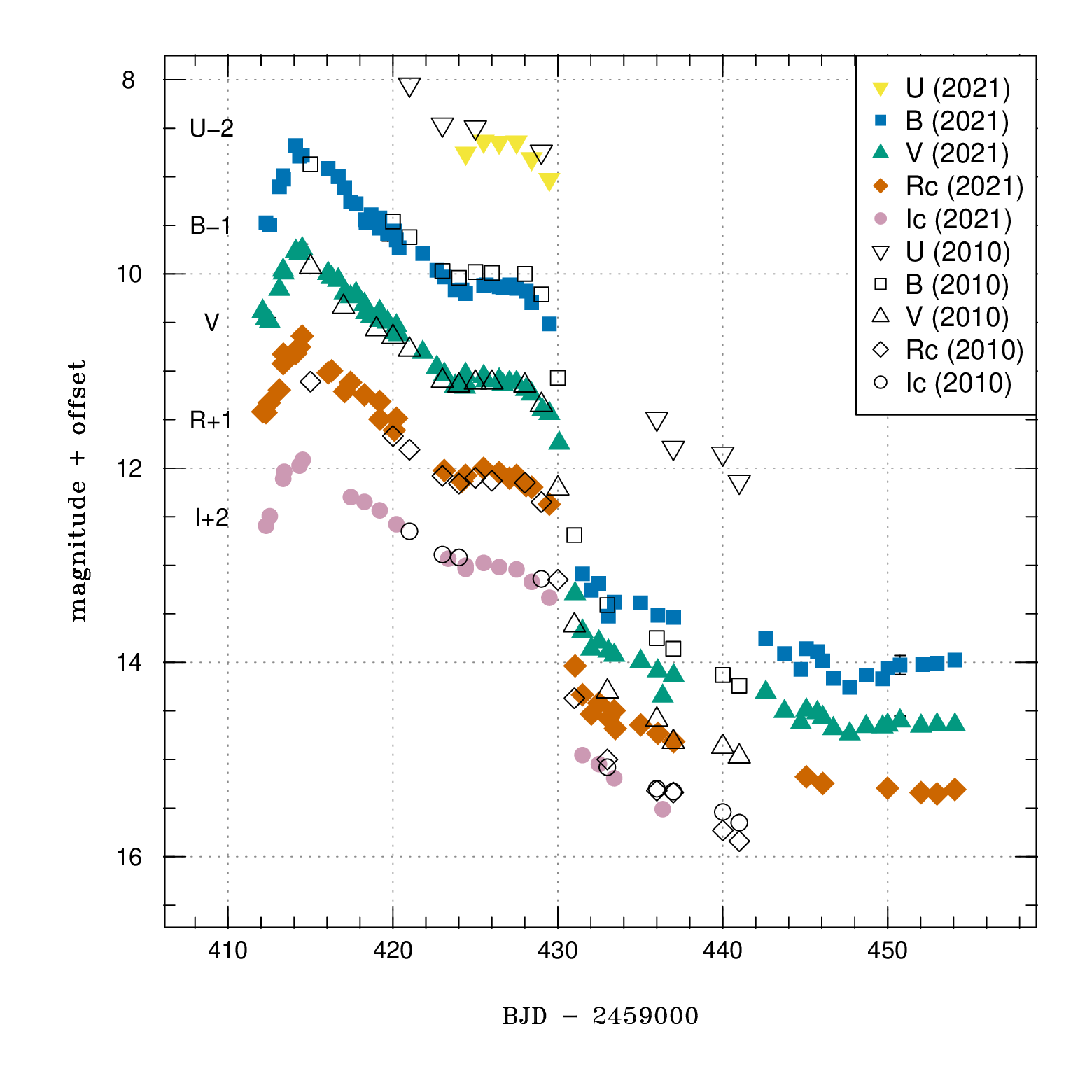}
  \includegraphics[width=80mm]{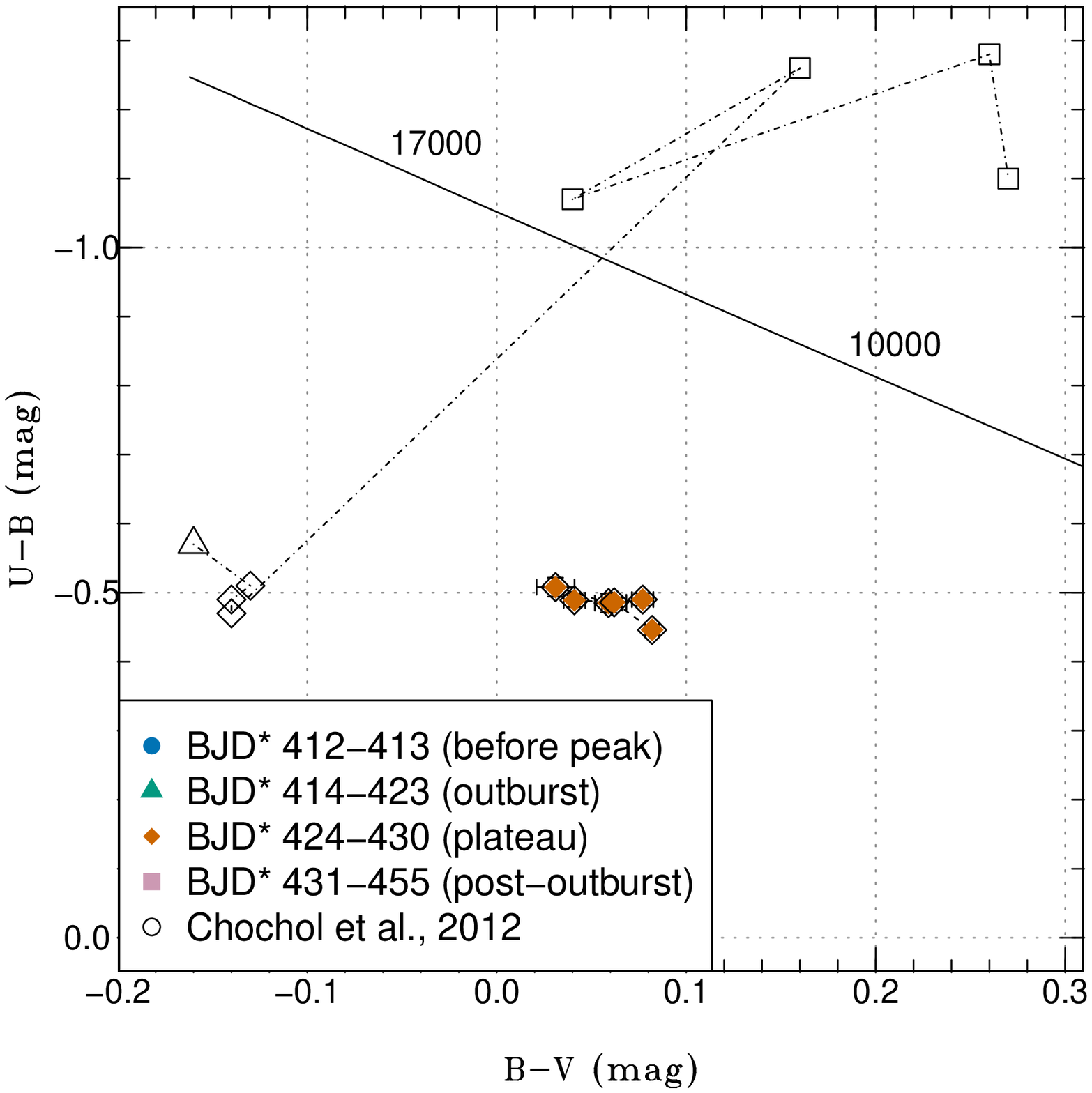}
  \includegraphics[width=80mm]{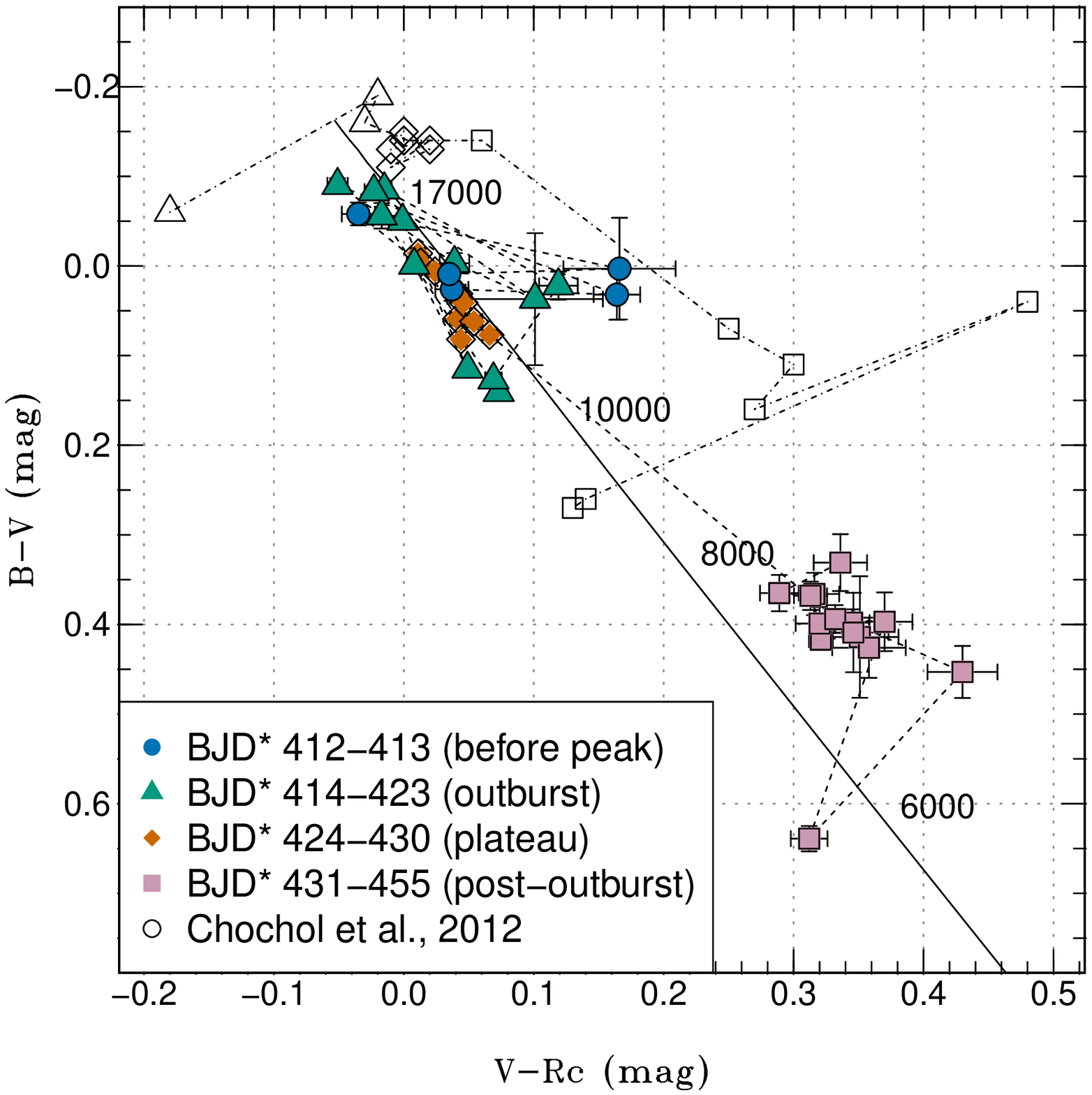}
  \includegraphics[width=80mm]{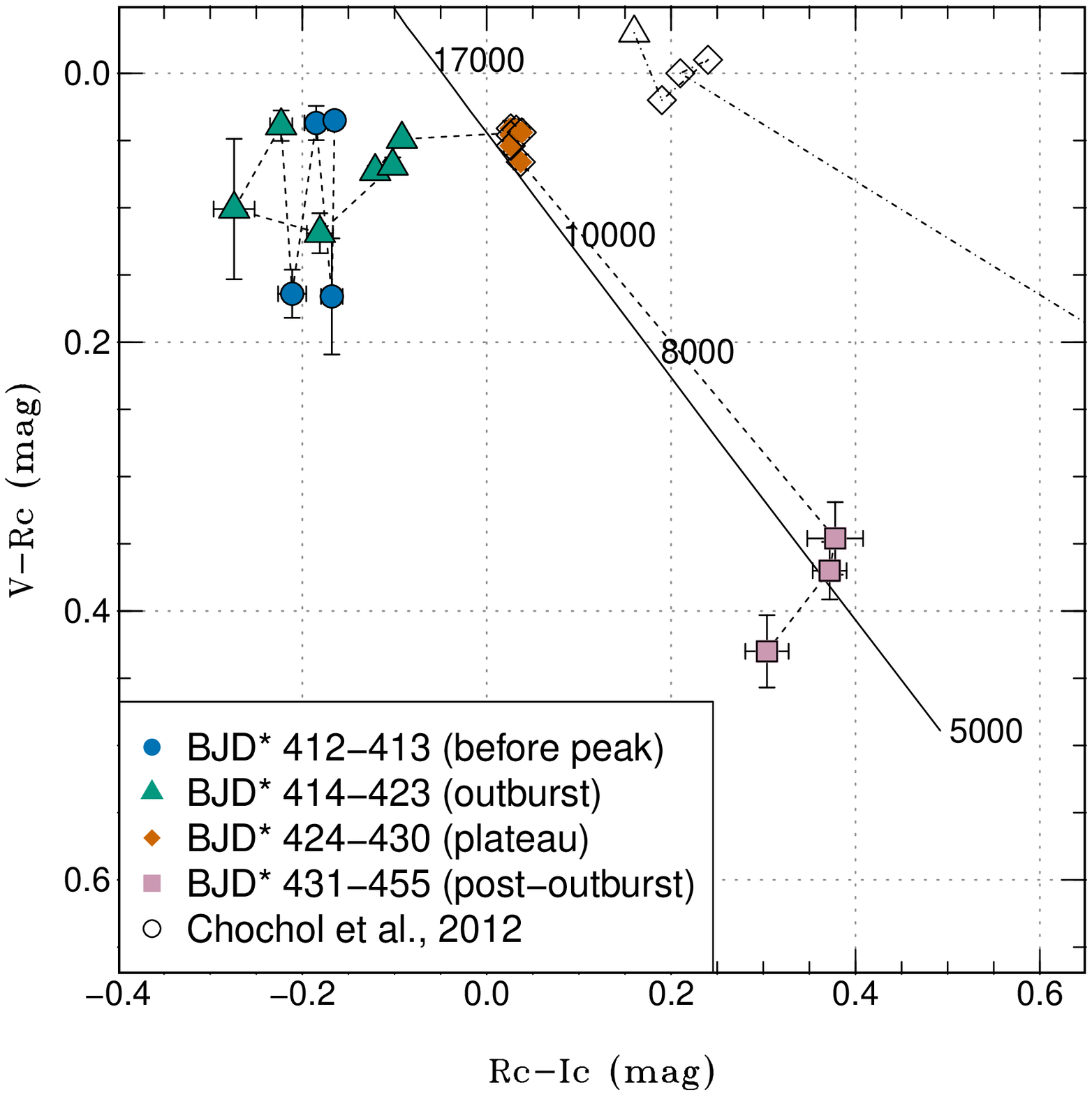}
 \end{center}
 \caption{
 Upper left panel: multi-color light curves in $U$ (yellow inverted triangles) $B$ (blue squares), $V$ (green triangles), $Rc$ (orange diamonds), and $Ic$ (pink circles) band. 
 {The light curves are vertically shifted for a better visualization.}
 Filled and open symbols represent the data of 2021 superoutburst (this paper) and the 2010 superoutburst \citep{cho12j2138}.
 Upper right panel:
 Color evolution of 2021 superoutburst of V627 Peg in the $U-B$ vs. $B-V$ plane.
 Blue circles: BJD 2459412-2459413 (before the peak of the outburst), green triangles: BJD 2459414-24594123 (outburst decline), orange diamonds:  BJD 2459424-2459430 (plateau phase), and pink squares:  BJD 2459431-2459455 (post outburst).
 Open symbols show the color evolution of V627 Peg during its 2010 superoutburst \citep{cho12j2138}.
 The shapes of open symbols have  the same meaning as the 2021 superoutburst.
 The solid line represents the color of blackbody emission, {and the corresponding temperature is indicated with the text.}
 Lower left panel: 
 Color evolution in the $B-V$ vs. $V-Rc$ plane.
 Symbols are same as in upper right panel.
 Lower right panel: 
 Color evolution in the $V-Rc$ vs. $Rc-Ic$ plane.
 Symbols are same as in upper right panel.}
 \label{fig:colorevol}
\end{figure*}

Based on our multi-color photometric observations, figure \ref{fig:colorevol} shows the color evolution of V627 Peg during its 2021 superoutburst (filled symbols) along with the  2010 superoutbursts (open symbols; \cite{cho12j2138}).
{In figure \ref{fig:colorevol}, we only plot the multi-color data from the Crimean Astrophysical Observatory, the Star\'a Lesn\'a Observatory, the Crimean laboratory of Sternberg Astronomical Institute, and the Observatory Sanglokh, which are calibrated with the same fashion as \citet{cho12j2138} and can be directly compared the data of \citet{cho12j2138} from its 2010 superoutburst.}
We note that in table 2 of \citet{cho12j2138}, the $Ic$-band magnitude was erroneously changed to +1.35 mag. 
However, the color-color diagrams in \citet{cho12j2138} were built without this addition. 
The authors (S.Shugarov) apologize to all readers for this mistake.
In the upper left panel of figure \ref{fig:colorevol}, the multi-color light curves in the $U, B, V, Rc$ and $Ic$ bands are presented.
The $U-B$ vs. $B-V$, $B-V$ vs. $V-Rc$ and $V-Rc$ vs. $Rc-Ic$ color-color diagrams are shown in the upper right, lower left and lower right panels, respectively.
The different symbols in the color-color diagrams represent the different stages of the superoutburst; blue circles: BJD 2459412-2459413 (before the peak of the superoutburst), green triangles: BJD 2459414-24594123 (outburst decline), orange diamonds:  BJD 2459424-2459430 (plateau phase), and pink squares:  BJD 2459431-2459455 (post superoutburst).
We also show the blackbody color (solid lines) and multi-color results in the 2010 superoutburst from \citet{cho12j2138} (open symbols) in the color-color diagrams.
We note that {observations made by different observers on the same night are plotted as separate points.}

The overall profile of the light curve of each band essentially traces the $V$ band light curve.
In the color-color diagrams, at the rise to the peak, V627 Peg is slightly redder in $B-V$, $V-Rc$ and $Rc-Ic$  color than  the optical peak.
This is  normal behavior in DN outbursts (e.g., \cite{bai80DNcolor}).
{The color temperature during the superoutburst was 13000-17000 K, within the range of that of other DN outbursts [e.g., \citet{shu21aylac}].}
The plateau phase showed almost the same color in $B-V$ and $V-Rc$ as in the outburst, while it was $\sim0.1$ mag redder in the $Rc-Ic$ color.
{The constant luminosity and gradual reddening in the plateau phase suggest that the disk was somehow in a quasi-steady state and the accretion rate is gradually decreasing, making the contribution from the outer disk greater.}
In the post-outburst phase, V627 Peg was redder than in the outburst and plateau phase in all $B-V$, $V-Rc$, and $Rc-Ic$ colors, {with the color temperature of $\sim 7000$ K.}
Although its color trend is consistent with other DN outbursts \citep{bai80DNcolor, shu21aylac}, we note that the exact color is uncertain because there is a nearby star with $\sim$15 mag (see Section \ref{sec:2}), and most of our observations after the rapid decline from the plateau phase are probably contaminated.

When comparing the 2021 superoutburst with the 2010 one, 
the $U, B,$ and $Ic$ light curves were slightly brighter in the 2010 superoutburst than in the 2021 superoutburst, although the $V$ and $Rc$ band light curve showed almost the same magnitude in both the 2010 and 2021 superoutburst.
This trend is also seen in the color-color diagram; throughout the superoutburst, the $B-V$ color was redder and the $Rc-Ic$ color was bluer in the 2021 superoutburst, while the $U-B$ and $V-Rc$ color showed almost the same color in the 2010 and 2021 superoutbursts.
Therefore, the emissivity from the outermost and innermost part of the accretion disk is likely lower in the 2021 superoutburst  than in the 2010 superoutburst.
These differences suggest that the accretion disk structure and the mass distribution were different in the 2010 and 2021 superoutbursts.

\section{Discussion}
\label{sec:4}

\subsection{Comparison between the three superoutbursts of V627 Peg}
\label{sec:4.1}

\begin{figure}[tbp]
 \begin{center}
  \includegraphics[width=80mm]{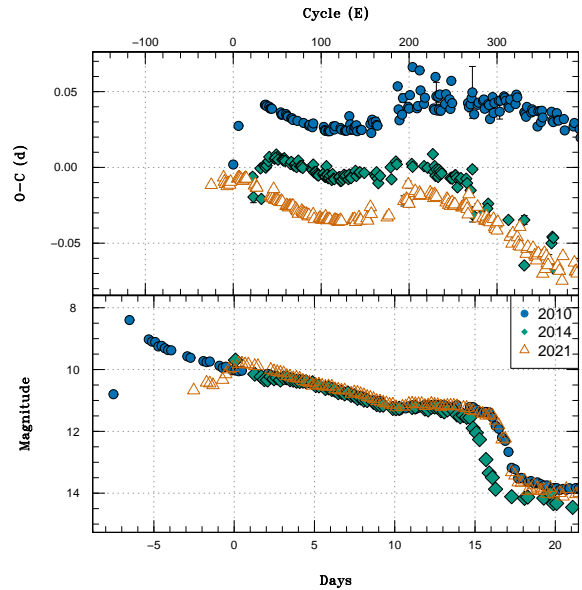}
 \end{center}
 \caption{$O-C$ diagram of superhump periods (upper panel) and light curves (lower panel) during the V627 Peg 2010 (blue circles), 2014 (green diamonds), and 2021  (orange triangles) superoutbursts.
 0.055120 d was used for $C$ \citep{Pdot}. 
 The Cycle of superhumps and Days are horizontally shifted so that the $O-C$ profiles during stage B superhumps are best matched.
 The $O-C$ diagram is vertically shifted for a visualising purpose.
 The light curves and $O-C$ data of the 2010 and 2014 superoutbursts were taken from \citet{pdot2, Pdot7}, respectively.}
 \label{fig:OCcomp} 
\end{figure}

In this section, we compare the three superoutbursts of V627 Peg, and explore whether its differences can be explained with the TTI model.
In figure \ref{fig:OCcomp}, we present the superhump period $O-C$ (upper panel) and the light curve (lower panel) during the V627 Peg 2010 (blue circle), 2014 (green diamond), and 2021 (orange triangle) superoutbursts.
We note that the datasets have been shifted horizontally so that the $O-C$ profiles during the initial part of the stage B superhumps are best matched.
Table \ref{tab:1} summarizes the parameters of three superoutbursts in V627 Peg.

\begin{table}
\caption{Observed parameters of three superoutbursts in V627 Peg.}
\centering
\label{tab:1}
\begin{tabular}{cccc}
  \hline     
     & \multicolumn{3}{c}{superoutburst} \\
     year  & 2010 & 2014 & 2021  \\
    \hline
    peak magnitude & 8.4 & 9.7 & 9.6 \\
    outburst duration (d) & 17.3 & 10.0 & 13.0\commentb \\
    plateau duration (d) & 5.7 & 4.8 & 6.0 \\
    precursor outburst & $\times$ & $\times$ & $\circ$ \\
    quiescence before outburst\commenta (yr) & ? & 4.47 & 6.73\\
  \hline
    \multicolumn{4}{l}{\commenta separation between outburst peaks.}\\
    \multicolumn{4}{l}{\commentb this value includes the precursor outburst.}\\
\end{tabular}
\end{table}

In the bottom panel of figure \ref{fig:OCcomp}, a large variety in the global light curve profiles of the three superoutbursts is recognized.
The 2010 superoutburst lasted longer for a few days and was 1-mag brighter at the peak than the 2014 and 2021 superoutbursts.
These features and the possible detection of early superhumps in the 2010 superoutburst \citep{pdot2} support 
the idea that the 2010 superoutburst was indeed a WZ Sge-type superoutburst accompanying an early superhump phase.
On the other hand, the fainter peak magnitude, the shorter outburst duration, and the detection of ordinary superhumps from the very early days of the outburst in the 2014 and 2021 superoutbursts, are consistent with SU UMa-type superoutbursts lacking an early superhump phase \citep{Pdot7}.
Furthermore, the 2021 superoutburst showed a clear precursor outburst, which was not observed in the 2010 superoutburst of V627 Peg and has never been observed in WZ Sge-type superoutbursts \citep{kat15wzsge}, indicating that this 2021 superoutburst was certainly a SU UMa-type superoutburst.
Even in a single object, the presence and profile of  precursor outbursts vary between the superoutbursts (e.g., \cite{uem05tvcrv}) and therefore the difference in the initial phase of the 2014 and 2021 superoutbursts is not  unique to V627 Peg.
Once the stage B superhumps appeared, these three superoutbursts showed a similar decline rate and entered the plateau phase around the same magnitude { of $V\sim 11.0$}.
This result emphasises that the 3:1 resonance and superhumps are the key mechanism of DN superoutbursts \citep{osa89suuma,osa03DNoutburst}.

The 2021 superoutburst showed fully developed superhumps on the rise to the superoutburst peak, whose periods are undoubtedly shorter than the one of stage A superhumps observed in the 2014 superoutburst (see figure \ref{fig:dailyPDM}). 
On the other hand, the 2014 superoutburst showed a clear stage A superhump phase even after it reached the peak magnitude.
Therefore, these results mean that the superhump in the 2021 superoutburst of V627 Peg started to evolve even during its precursor phase.
While in most cases of SU UMa-type superoutbursts, the stage A-B transition of superhumps occurs after the superoutburst peak in the well-observed systems (e.g., \cite{kat13j1939v585lyrv516lyr}), some superoutbursts show superhumps even during a precursor outburst
\citep{kat97tleo, uem05tvcrv, ima09j0532, osa14v1504cygv344lyrpaper3, ima17qzvir}.
Based on the relationship between the mass ratio and duration of stage A superhumps \citep{kat15wzsge}, and applying a mass ratio of 0.1, the expected duration of stage A superhumps is around 20 cycles, which is equal to $\sim$ 1.1 d with the orbital period of 0.0545 d.
This value is consistent with the length of stage A superhumps in the 2014 superoutburst.
In the case of the 2021 superoutburst, the developed superhumps are observed on BJD 2459413.2 during the rise to the superoutburst peak, indicating that the 3:1 resonance started to work as early as BJD 2459412.
This epoch corresponds to around the maximum of the precursor outburst.
Thus, the maximum disk radius during the precursor outburst should be beyond the 3:1 resonance radius, and superhumps can develop during the decay in the precursor outburst.
Our data during this phase showed small variations, but the quality of the data was insufficient to determine their periodicity.

The precursor outburst was observed in the 2021 superoutburst of V627 Peg, while it was not observed in the 2010 and 2014 superoutburst \citep{pdot2, Pdot7}.
According to \citet{osa03DNoutburst}, a precursor outburst of SU UMa-type superoutbursts can be observed if the disk radius does not reach the tidal truncation radius \citep{pac77ADmodel} at the onset of an outburst, while it is still larger than the 3:1 resonance radius.
Therefore, in the TTI model, the precursor in the 2021 superoutburst indicates that the maximum disk radius of the 2021 superoutburst was smaller than the tidal truncation radius.
On the other hand, the 2014 superoutburst probably reached the tidal truncation radius at the peak, although the disk mass was likely too small to reach the 2:1 resonance radius, which is located beyond the tidal resonance radius for its large mass ratio \citep{osa02wzsgehump, wak17asassn16eg}.
This result is consistent with the discussion above, that the 3:1 resonance became active around the peak of the precursor outburst in the 2021 superoutburst.

As shown in the top panel of figure \ref{fig:OCcomp}, the duration of stage B superoutbursts also differs between the three superoutbursts.
The 2010 superoutburst, featuring early superhumps, showed the longest stage B superhumps, continuing even after V627 Peg entered the plateau phase.
The 2014 and 2021 superoutbursts without an early superhump phase had almost the same duration.
A similar phenomenon was observed in V3101 Cyg, which showed SU UMa-type superoutbursts during the rebrightening phase after the main WZ Sge-type superoutburst in 2019 \citep{tam20j2104, ham21V3101Cygrebrightening}.
In the case of V3101 Cyg, the duration of stage B superhumps in the main superoutburst lasted for 10 days, whereas in the two rebrightening superoutbursts lacking early superhumps, the duration of stage B superhumps was just 7 days \citep{tam20j2104}.
These differences in the duration of stage B superhumps in V627 Peg and V3101 Cyg could be attributed to the presence of the early superhump stage in a respect superoutburst.
If the 2:1 resonance occurs in a superoutburst (WZ Sge-type superoutburst), the disk condition when the 3:1 resonance starts to work may be different from those in a SU UMa-type superoutburst.
In a WZ Sge-type superoutburst, an effective tidal dissipation by the 2:1 resonance will push the disk matter inside \citep{lin79lowqdisk, osa02wzsgehump}.
Therefore, the inner disk would become more massive before the 3:1 resonance becomes active than in an accretion disk that does not experience the 2:1 resonance.
Since stage B in the superhump evolution is regarded as the inward progression of the eccentricity from the 3:1 resonance radius (e.g., \cite{Pdot, nii21a18ey}),  a more massive inner disk will result in a slower penetration of the eccentricity.
This effect can result in more time for the disk to become fully eccentric, and hence a longer duration of stage B superhumps can be observed in superoutbursts accompanied by early superhumps. 
This interpretation is also consistent with the differences in the color evolution between the 2010 and 2021 superoutbursts.
The 2010 superoutburst was brighter than the 2021 superoutburst in the $U, B$ and $Ic$ bands, suggesting that the emissivity from the innermost and outermost part of the disk was higher in the 2010 superoutburst.
A disk experiencing the 2:1 resonance might have more mass in the inner disk, yielding bright bluer bands, 
and the removal of angular momentum by the 2:1 resonance can also expand the disk outward; a low-temperature outer disk could be observed as a brighter $Ic$-band flux in the 2010 superoutburst.

The duration of the plateau phase is the shortest in the 2014 superoutburst, while the 2010 and 2021 superoutbursts showed almost the same duration.
This trend is different from the light curve profiles and the superhump evolution among the three superoutbursts in V627 Peg.
Since the superhump evolution in the 2010 and 2021 superoutbursts was greatly different, the nature of this plateau phase is not likely related to the superhumps and therefore the occurrence of the resonance.
\citet{Pdot} proposed that this plateau phase is a kind of rebrightening (type A rebrightening; \cite{ima06tss0222}), which is a subsequent outburst after the main superoutbursts; it is observed in most WZ Sge-type DNe and some SU UMa-type DNe.
As the profiles of the rebrightenings are believed to be inherent to the object regardless of the presence of early superhumps (e.g., \cite{kim16alcom, kim21EGCnc}), the profile differences during the plateau phase between three V627 Peg superoutbursts indicates that this phase may not be a similar phenomenon of a rebrightening.
Moreover, the three superoutbursts of V627 Peg did not show any dip between the superoutburst and the plateau phase, whereas type A rebrightenings are observed a few  days after the rapid decline at the end of the superoutburst.
Even though the physical mechanism of rebrightenings are not yet well understood, in the most models of rebrightening, the mechanism is explained by another series of outbursts from a disk which once has dropped into a cool state \citep{kat98super, ham00DNirradiation, osa01egcnc, mey15suumareb}.
{These models are inconsistent with our result, as the 2021 superoutburst showed a continuous transition from the superoutburst to the plateau phase, rather than another series of outbursts.}
The origin of these differences among the three superoutbursts may be related to the duration of quiescence before these three superoutbursts.
The 2014 superoutburst was observed only four years later the previous one, whereas the 2021 superoutburst was observed seven years after the previous one.
Taking into account that the mass transfer rate of DNe is regarded as constant, a longer duration of quiescence will likely result in a more massive disk and hence a longer duration in the plateau phase.

\subsection{implication on the outburst mechanism of WZ Sge-type dwarf novae}
\label{sec:4.2}

As discussed in the previous subsection, the presence of a precursor outburst in the 2021 superoutburst indicates that the maximum disk radius was smaller in the 2021 superoutburst than in the 2014 superoutburst.
However, considering the duration of quiescence and a constant mass transfer rate in DNe, the initial disk mass at the onset of the outburst should be larger in the 2021 outburst than in the 2014 outburst.
Therefore, if both superoutbursts were triggered in the same mechanism and with the constant mass-transfer rate, the 2021 superoutburst should reach a larger maximum disk radius, which does not agree with our observations and the interpretation of precursor outburst in the TTI model.

One possible scenario to understand the 2021 superoutburst {in the TTI model} is to decrease the mass transfer rate over the long term, resulting in a less-massive accretion disk at the onset of the 2021 superoutburst compared to that of the 2014 superoutburst.
Since the quiescence length before the 2021 superoutburst is about twice of the 2014 superoutburst, the mass transfer rate must at least decrease by a factor of two.
A long-term change of the mass transfer rate is considered in some type of CVs (e.g., Z Cam-type DNe and VY Scl-type CVs; \cite{war95book}), and the angular momentum loss of these systems is by magnetic braking (\cite{kni11CVdonor} and reference therein).
However, since V627 Peg is a CV below the period-gap, the dominant mechanism for angular momentum loss should be gravitational wave radiation, and therefore a sudden change in the angular momentum loss is less expected.
Furthermore, since the total outburst duration of the 2021 superoutburst is longer than that of the 2014 superoutburst, the initial disk mass is expected to be more massive in the 2021 superoutburst.

Another scenario is that the 2021 superoutburst might be an inside-out outburst, while the 2010 and 2014 superoutbursts were  triggered as an outside-in outburst.
This is because, as studied by \citet{sma84DI}, the inside-out outburst is initially triggered at the inner part of the disk; 
therefore, the maximum disk radius of inside-out outbursts will be smaller than that of outside-in outbursts, even if the initial disk mass is the same.
Although the observed rise timescale after the detection of the 2021 superoutburst is comparable to other WZ Sge-type DNe \citep{ish02wzsgeletter, waa07gwlibiauc, tem07cbet922gwlib}, the contamination of the quiescent data prevents us from measuring the exact rise rate from quiescence.
The indication of inside-out outbursts in WZ Sge-type DNe may lead one to revise the idea that WZ Sge-type DNe should be triggered as outside-in outbursts only based on their fast rise to the outburst \citep{ish02wzsgeletter, Pdot}.
A low value of the viscosity parameter $\alpha$ \citep{sma93wzsge, osa95wzsge} or the truncation of the inner disk by a magnetic WD \citep{war96wzsge, mat07wzsgepropeller, kuu11wzsge} are common ideas to explain the absence of inside-out outbursts in WZ Sge-type DNe.
If the low $\alpha$ explanation is correct, one might conclude that V627 Peg has a moderate value for the viscosity parameter $\alpha$, which would enable both inside-out and outside-in outbursts.
Since the mass ratio of V627 Peg is one of the largest value among WZ Sge-type DNe, the viscosity parameter $\alpha$ is not so small compared with the other WZ Sge-type DNe.
On the other hand, if magnetic truncation were the crucial factor, the primary WD of V627 Peg would be more weakly magnetized than the WDs in other WZ Sge-type DNe. 

In EG Cnc and AL Com, \citet{kim16alcom, kim21EGCnc} suggested that the initial disk mass rules whether a particular superoutburst is observed as a WZ Sge-type or SU UMa-type superoutburst.
In addition to this fact, the large variety in the three superoutbursts of V627 Peg suggests that there are more factors which determine the outburst type and profile in WZ Sge-type DNe.
These results emphasize the importance of continuous observations and monitoring of known WZ Sge-type DNe.

\section{Summary}
\label{sec:5}

We report photometric observations during the superoutburst of the WZ Sge-type DN V627 Peg in 2021.
Our key findings are summarized as follows:

\begin{itemize}

\item 
The 2021 superoutburst of V627 Peg showed a precursor outburst, which is quite uncommon in WZ Sge-type DNe.
The peak magnitude of the 2021 superoutburst was 1-mag fainter than that of its 2010 superoutburst, which showed probable early superhumps.
After the precursor outburst, well-developed ordinary superhumps were observed even on the rise to the optical peak,
suggesting that this 2021 superoutburst did not feature an early superhump phase.
Thus, the 2021 superoutburst of V627 Peg lacked an early superhump phase, even though it was a superoutburst from a WZ Sge type DN.

\item 
The 2021 superoutburst showed a plateau phase for about six days, similar to that in its 2010 superoutburst.
The plateau phase showed $B-V$ and $V-Rc$ colors similar to those of the superoutburst decline, while the $Rc-Ic$ color was $\sim 0.1$ mag redder than the outburst.

\item 
Unlike the superhump features of three V627 Peg superoutbursts, the plateau phase was shortest in the 2014 superoutburst.
Our results suggest that the mechanism of the plateau phase might not be identical to rebrightenings, which is often observed in WZ Sge-type DNe.
One possible explanation is the longer quiescence period before the 2021 superoutburst, compared to the 2014 superoutburst, which may make the disk before the outburst more massive and the plateau phase longer.

\item 
The duration of state B superhumps was observed to be longer in the 2010 superoutburst than in the 2014 and 2021 superoutbursts.
Once the 2:1 resonance takes place in the accretion disk, more gas will accumulate in the inner part of the accretion disk.
Since the duration of stage B superhumps is regarded as the progression of eccentricity from the 3:1 resonance radius to the inner part of the disk, a more massive inner disk may result in a longer duration of stage B superhumps, as observed in the 2010 superoutburst.

\item
Taking into account that the 2021 superoutburst showed a precursor outburst, we expect a smaller maximum disk radius in the outburst peak than in the 2014 superoutburst.
However, the duration of quiescence was longer before the 2021 superoutburst than the 2014 superoutburst.
This result can be explained {in the thermal-tidal instability model} if the 2021 superoutburst was triggered as an inside-out outburst, or if the long-term mass transfer rate decreased by a factor of two.
These results suggest that the quiescence disk structure and the outburst mechanism in WZ Sge-type DNe may have a larger variety than previously expected.

\end{itemize}

\begin{ack}

Y. T. acknowledges support from the Japan Society for the Promotion of Science (JSPS) KAKENHI Grant Number 21J22351. 
T. K and D. N. acknowledge support from the Japan Society for the Promotion of Science (JSPS) KAKENHI Grant Number 21K03616.
Part of this work was supported by the Slovak Research and Development Agency under contract No. APVV-20-0148 (S. Y. S., P. A. D., and T. M.) and Grant VEGA 2/0030/21 (S.Y.S.).
K. M. acknowledges support from the Japan Society for the Promotion of Science (JSPS) KAKENHI Grant Number 19K03930.
O. A. acknowledges the partial support from the Ministry of Science and Higher Education of the Russian Federation (grant 075-15-2020-780).
V. K. acknowledges the Scholarship of the Slovak Academic Information Agency SAIA and Dr. \v{S}tefan Parimucha, Assoc. prof. at the Faculty of Science, P.J. \v{S}af\'{a}rik University in Ko\v{s}ice.

The authors acknowledge amateur and professional astronomers around the world who have provided the data of V627 Peg with the VSNET collaboration.

Lasair is supported by the UKRI Science and Technology Facilities Council and is a collaboration between the University of Edinburgh (grant ST/N002512/1) and Queen’s University Belfast (grant ST/N002520/1) within the LSST:UK Science Consortium. ZTF is supported by National Science Foundation grant AST-1440341 and a collaboration including Caltech, IPAC, the Weizmann Institute for Science, the Oskar Klein Center at Stockholm University, the University of Maryland, the University of Washington, Deutsches Elektronen-Synchrotron and Humboldt University, Los Alamos National Laboratories, the TANGO Consortium of Taiwan, the University of Wisconsin at Milwaukee, and Lawrence Berkeley National Laboratories. Operations are conducted by COO, IPAC, and UW. This research has made use of ``Aladin sky atlas'' developed at CDS, Strasbourg Observatory, France \cite{Aladin2000,Aladinlite}.

This work has made use of data from the European Space Agency (ESA) mission
{\it Gaia} ({https://www.cosmos.esa.int/gaia}), processed by the {\it Gaia}
Data Processing and Analysis Consortium (DPAC,
{https://www.cosmos.esa.int/web/gaia/dpac/consortium}). Funding for the DPAC
has been provided by national institutions, in particular the institutions
participating in the {\it Gaia} Multilateral Agreement.

\end{ack}

\section*{Supporting Information}
The following Supporting Information is available on the online version of this article: table E1, table E2 and table E3.


\appendix

\setcounter{figure}{0}
\renewcommand{\thefigure}{A\arabic{figure}}

\section{On the thermal-tidal instability model}

The following sections of the appendix were added to clarify the claims in the referee report from the anonymous referee.
We appreciate the referee's comments which improved our understanding of the TTI model.
The main comments of the referee report were based on \citet{neu16htcas, neu20HTCasoutburstdopmap, ama21ezlyn} and they observationally questioned that an accretion disk in quiescence of SU UMa-type DNe can be larger than the 3:1 resonance radius, which is not expected in the current TTI model.

\subsection{treatment of disk radius}

\citet{neu16htcas, ama21ezlyn} converted the double-peaked profile of the phase-averaged emission line of H$\alpha$ to the disk radius assuming that this profile originates from the Keplerian disk around the primary white dwarf.
The estimated disk radius in quiescence is larger than the 3:1 resonance radius in both the cases of HT Cas and EZ Lyn, and they suggested that the quiescent disk can extend beyond the 3:1 resonance radius.
This is significantly larger than the disk radius estimated through the eclipse observations in HT Cas (0.2-0.3$a$, where $a$ is a binary separation; e.g., \cite{vri02htcas}), which is usually considered the most reliable measurement way for a disk radius.

Generally speaking, it is known that an estimated radius from peak separations of  emission lines gives a systematically larger value than the disk radius measured by eclipse observations (e .g., \cite{wad88zcha, mar88ippeg, war95book}). 
{\citet{neu16htcas, ama21ezlyn} argued that the accretion disk radius from the peak separation of emission lines is larger than the radius given by eclipse analysis, but still is truncated at the tidal truncation radius.
Meanwhile, based on the observation of peak separations, \citet{ech07ugem} estimated that the disk radius of U Gem in quiescence becomes as large as the Roche lobe radius, and the hot spot is located around the L1 point.  
\citet{ham20CVreview} pointed out that "such a large disc would in any case pose severe problems from a dynamical point of view" in this result by \citet{ech07ugem}.}

According to the numerical simulation of WZ Sge in quiescence by \citet{mey98wzsge}, an accretion disk can reach the 3:1 resonance in quiescence if the disk is not inviscid. 
However, the density of the outer disk remains rather low to trigger an outburst and unlikely to govern the dynamics of the disk as a whole. 
Although the above statement is based on  WZ Sge
and there might be an objection that the case is different from
a SU UMa-type DN, what \citet{neu16htcas, ama21ezlyn} have observed can be a similar situation.
Along with these papers, \citet{ech21ahher} also proposed that some gas is located outside the truncation radius of AH Her on the basis of the Doppler map.

In this paper, we regarded the disk radius as the one obtained through the eclipse observations. 
This is mainly because the quiescence of SU UMa-type DNe does not show any positive superhumps except for a period between a precursor outburst and a superoutburst (e.g., \cite{osa13v344lyrv1504cyg, kat13j1939v585lyrv516lyr});
therefore, the 3:1 resonance is not induced in the quiescence. 
This respect strongly suggests that even though there can be gas outside of the 3:1 resonance radius as inferred by the spectroscopic observations, the representative disk radius regulating the behavior of SU UMa-type DNe in the TTI model is the disk radius provided by the eclipse observations, rather than the spectroscopically calculated radius.

\subsection{Jacobian in conversion from Doppler tomography to spacial coordinate}

Assuming that an accretion disk follows the Kepler rotation around the primary white dwarf, one can convert a Doppler map with velocity coordinate to an image with spatial coordinate.
We note that since this conversion assumes a Keplerian disk, some components which do not follow the Kepler motion in the binary (e.g., secondary Roche lobe, mass stream from the L1 point) would not be converted correctly.
Furthermore, the Doppler map in velocity coordinates can be easily interpreted as an observed spectrum by projecting the map to the viewing angle of an observer.
The spatially converted image does not have this advantage.
With these attentions in mind, in this conversion, one needs to multiply the suitable Jacobian, which is shown in equation \ref{eqa1}, where $v_x$ and $v_y$ is the velocity coordinate in Doppler tomography, $x$, and $y$ is the spatial coordinate, $G$ is the gravitational constant, $M$ is the primary mass, and $r~(=\sqrt{x^2+y^2})$ is the radius of the disk.

\begin{equation}
\label{eqa1}
    J = 
\begin{vmatrix}
\frac{dv_x}{dx} & \frac{dv_x}{dy} \\
\frac{dv_y}{dx} & \frac{dv_y}{dy} \\
\end{vmatrix}
= 2GMr^{-3}
\end{equation}

\begin{figure*}[tbp]
 \begin{center}
  \includegraphics[width=60mm]{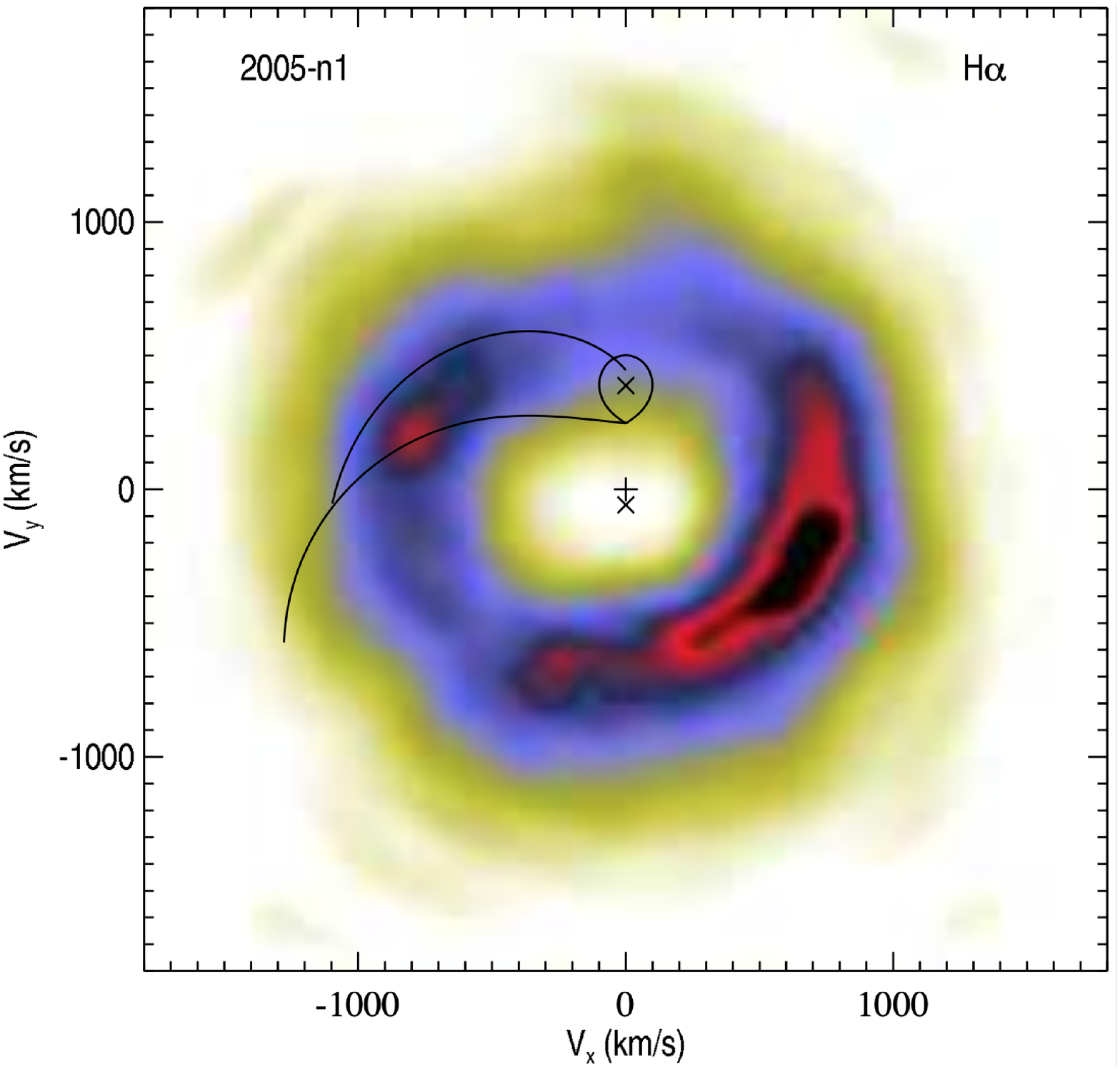}
  \includegraphics[width=60mm]{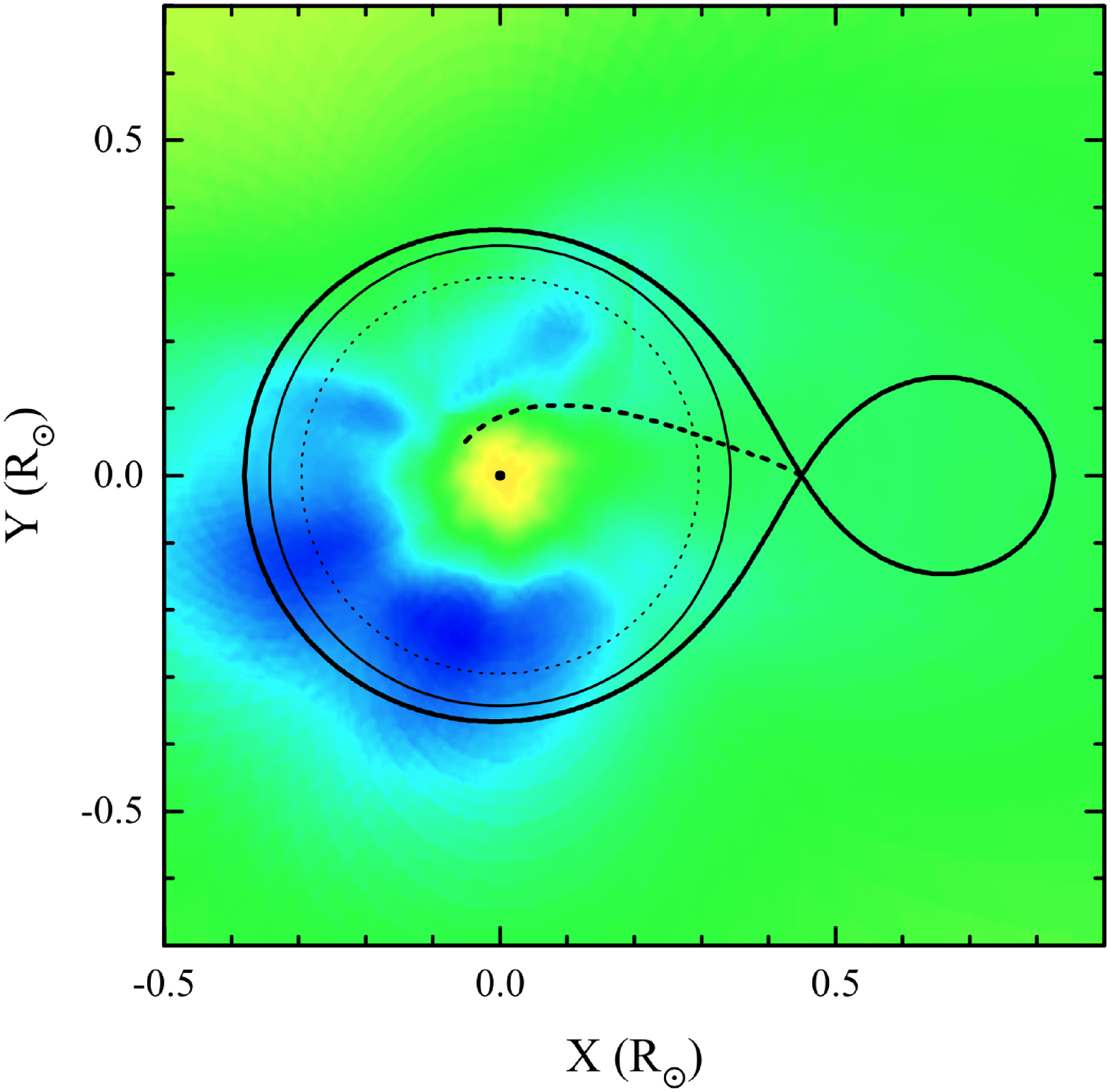}
  \includegraphics[width=60mm]{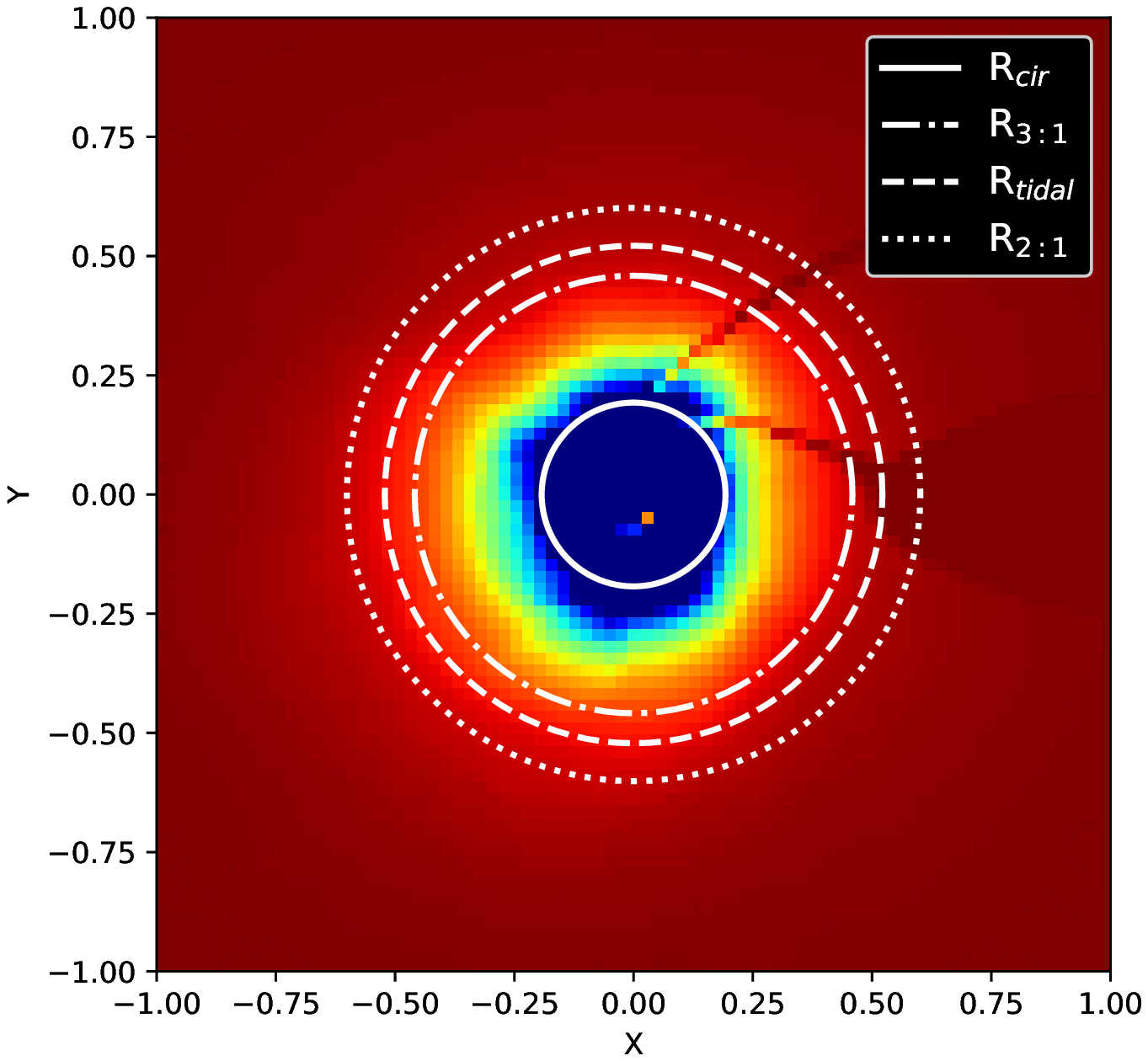}
  \includegraphics[width=60mm]{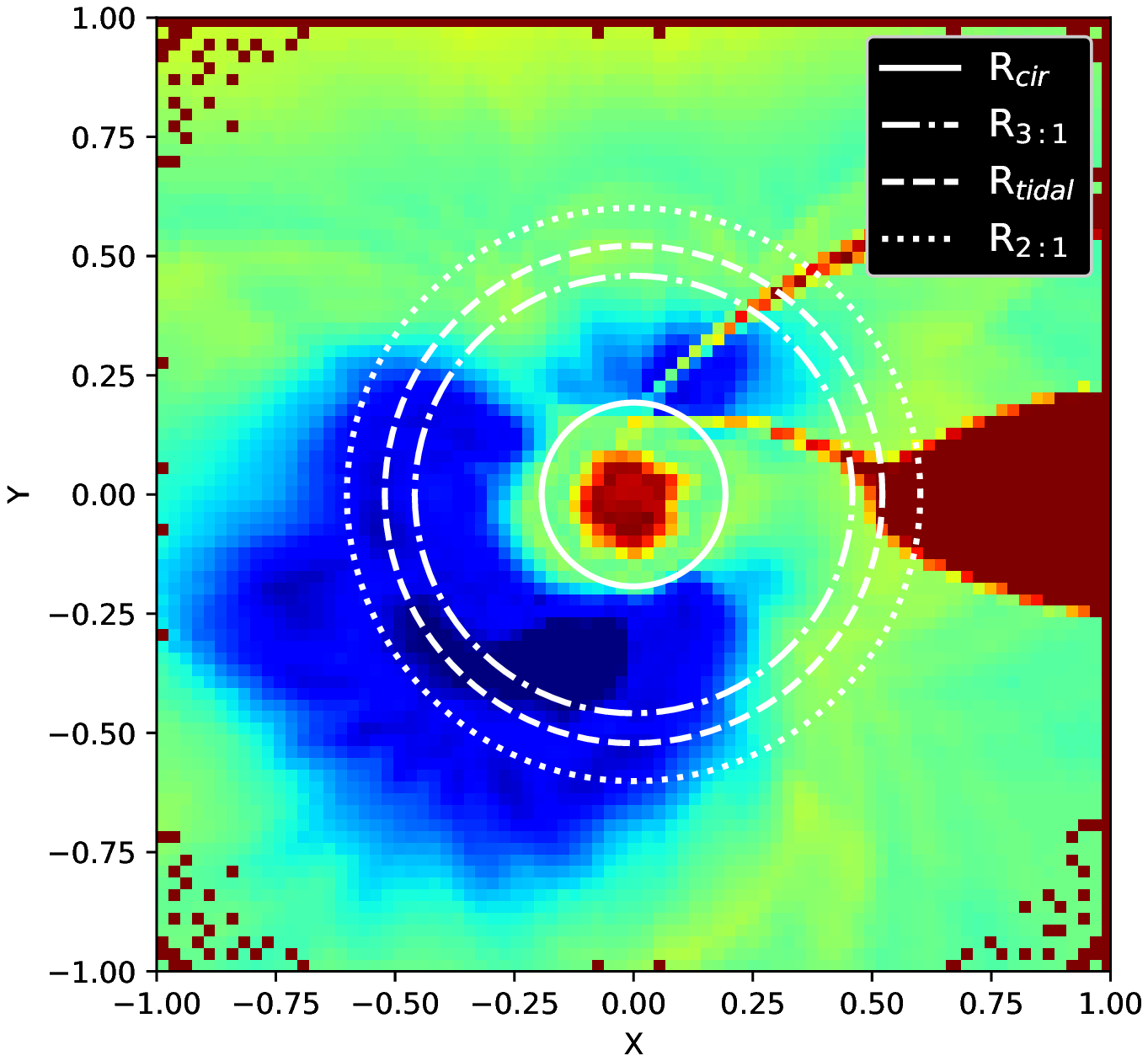}
 \end{center}
 \caption{Copied Doppler map of H$\alpha$ from figure 6 of \citet{neu16htcas} (upper left-hand panel) and spatially converted image from figure 10 of \citet{neu16htcas} (upper right-hand panel). 
 We also show the converted images with our own code in spatial coordinate with Jacobian (lower left-hand panel), and without Jacobian (lower right-hand panel).
 Lower panels are shown with spatial coordinates normalized by the binary separation.
 The primary white dwarf and the secondary is located at $(X,Y)=(0,0)$ and $(X,Y)=(1,0)$, respectively.
 The solid, dot-dashed, dashed, and dotted lines in the lower panels correspond to the circularization radius, the 3:1 resonance radius, the tidal truncation radius, and the 2:1 resonance radius {from inner to outer circles} applying the same binary parameter of HT Cas as \citet{neu16htcas}.
 We note that since the conversion was performed from the published image in \citet{neu16htcas}, 
 one sees some artificial strictures in the lower panels (e.g., secondary Roche lobe, gas stream from L1 point.)}
 \label{fig:htcas}
\end{figure*}

Figure 10 of \citet{neu16htcas} or figure 14 of \citet{ama21ezlyn} does not likely take into the Jacobian for their conversion from Doppler tomography to spatial XY coordinates.
Since Jacobian has a strong dependence on the radius  ($\propto r^{-3}$), an omission of Jacobian gives a biased impression that the emission from the outer disk much exceeds that from the inner disk. 
We note that the same issue has already been pointed out by \citet{kat21CzeV404} in figure 12 of \citet{kar21CzeV404}.
In figure \ref{fig:htcas}, we show the copied Doppler map of H$\alpha$ of \citet{neu16htcas} using data observed in 2005 (upper left-hand panel). 
The converted images in spatial coordinates without taking into Jacobian (lower {right-hand} panel of figure \ref{fig:htcas}) and with Jacobian (lower {left-hand} panel of figure \ref{fig:htcas}) are also presented. 
The converted image without Jacobian of figure \ref{fig:htcas} very closely resembles figure 10 of \citet{neu16htcas} or the upper right-hand panel of figure \ref{fig:htcas}. which is presented as a converted spatial map in \citet{neu16htcas}. 

\subsection{double-humped profile in quiescence of WZ Sge-type dwarf novae}

\citet{ama21ezlyn} showed that the double-humped orbital profile of EZ Lyn in quiescence can be explained with a spiral arm structure in an accretion disk reaching the 2:1 resonance radius.
The double-humped pattern in the quiescence light curve has been observed in many WZ Sge-type DNe (e.g., WZ Sge, AL Com, V455 And: \cite{pat96alcom, ara05v455and}). 
Orbital variations in quiescence of DNe are often explained by an optically thin disk plus a hot spot, and the same fashion has also been taken into WZ Sge-type DNe (e.g. \cite{ski00wzsge}).  
\citet{kon15wzsgequihump}, on the other hand, showed that the quiescence profile of V455 And can be modeled with an accretion disk with four shock waves and with a disk radius smaller than 0.4$a$ (where $a$ is the binary separation).
Therefore, even if the quiescent light curve of EZ Lyn is modeled with the spiral arm pattern in a disk reaching the 2:1 resonance, it cannot be a unique explanation for the quiescent orbital profile of WZ Sge-type DNe.
Along with this point, since a spiral arm pattern is observed in various DNe, including objects above the period gap (e.g. \cite{har99ippeg, ste01spiralwave, gro01ugemspiral}), 
their result does not necessarily mean that the spiral arm structure must be triggered by the 2:1 resonance and the disk expands beyond the 2:1 resonance radius. 

Moreover, 
in AL Com, long-period superhump is observed in quiescence \citep{pat96alcom}, which can be explained with the precessing accretion disk reaching the 2:1 resonance radius \citep{kat13qfromstageA}.
This phenomenon provides evidence that even if a disk in quiescence reaches the 2:1 resonance radius, a spiral arm structure fixed with the binary rotational frame is not formed.


\bibliographystyle{pasjtest1}
\bibliography{cvs}


\end{document}